\documentclass[doublecol]{epl2} 

\usepackage{amssymb}
\usepackage{amsmath}
\usepackage{dsfont}
\usepackage{bm,bbm}
\usepackage{lipsum}
\usepackage{mathtools, cuted}
\usepackage{breqn}
\usepackage{xcolor}
\usepackage{soul}

\usepackage{amsthm}
\usepackage{verbatim}
\usepackage{hyperref}
\usepackage{url}

\title{CompLex: legal systems through the lens of complexity science.}
\shorttitle{CompLex: legal systems through the lens of complexity science.} 

\author{Pierpaolo Vivo\inst{1}, Daniel M. Katz\inst{2,3,4} \and J.B. Ruhl\inst{5}}

\shortauthor{P. Vivo \etal}

\institute{                    
  \inst{1} Quantitative and Digital Law Lab, Department of Mathematics, King’s College London, UK\\
  \inst{2} Illinois Institute of Technology, Chicago-Kent College of Law, Chicago, IL, USA\\
\inst{3} CodeX, The Stanford Center for Legal Informatics, Stanford, CA, USA\\
\inst{4} Center for Legal Technology \& Data Science, Bucerius Law School,  Hamburg, Germany\\
\inst{5} Vanderbilt University Law School, Nashville, TN, USA
}

\pacs{}{}

\abstract{
While ``complexity science'' has achieved significant successes in several interdisciplinary fields such as economics and biology, it is only a very recent observation that legal systems -- from the way legal texts are drafted and connected to the rest of the corpus, up to the level of how judges and courts reach decisions under a variety of conflicting inputs -- share several features with standard Complex Adaptive Systems. This review is meant as a gentle introduction to the use of quantitative tools and techniques of complexity science to describe, analyse, and tame the complex web of human interactions that the Law is supposed to regulate. We offer an overview of the main directions of research undertaken so far as well as an outlook for future research, and we argue that statistical physicists and complexity scientists should not ignore the opportunities offered by the cross-fertilisation between legal scholarship and complex-systems modelling.}

\begin{document}

\maketitle

\section{Introduction}
``Complexity science'' \cite{newman2011resource, gell2002complexity, miller2009complex, boccara2010modeling} -- a relatively recent incarnation of Statistical Physics -- has achieved significant successes in characterising diverse systems such as economic markets \cite{battiston2016complexity, farmer2009economy, hidalgo2009building}, ecological and biological systems \cite{sole2022ecological, holling2001understanding,cavagna2010scale} and the structure and dynamics of human societies \cite{bettencourt2010unified, gonzalez2008understanding, bettencourt2007growth}.  Such efforts vary but typically involve both formal models and the empirical evaluation of common trends of behaviour, resilience and critical points as well as multi-scale dynamic interaction between various system components.  

 Although a one-size-fits-all definition of what a \emph{complex system} is still escapes us, there is a general consensus that (i) it is generally formed of a large number of elementary constituents, (ii) such constituents interact with each other, typically by \emph{local} (short-range) rules, (iii) some of the macroscopic (collective) features and behaviour should emerge at larger scale without being hard-coded in the local rules, and (iv) there is an element of \emph{randomness}, which makes a probabilistic description usually more suitable than a deterministic one. 

Legal theorists, working largely in qualitative terms,\footnote{Asked to define the complexity of the Law more precisely, a legal scholar famously suggested ``I know it when I read it'' \cite{kades}.} have highlighted some ``complex systems''-type properties present in legal systems \cite{luhmann2004law, ruhl1995complexity, ruhl2015measuring, allen2012complexity, hathaway2000path} including the potential for \textit{emergent} system level behaviour \cite{ruhl2015measuring}. While physicists and applied mathematicians have expanded their traditional mandate to consider a range of adjacent topics such as social physics and econophysics \cite{perc2019social, perc2017statistical, ball2012society, castellano2009statistical, mantegna1999introduction}, less technical attention has historically been paid to modeling legal systems within the complex systems paradigm.  Indeed, it was only very recently that scholars have attempted to more rigorously characterise the underlying dynamics using modern quantitative and mathematically oriented methods \cite{vivo2024complexity, vivo2022physics, katz2020complex, leibon2018bending, ruhl2017harnessing, bommarito2017measuring, koniaris2018network, katz2010hustle, bourcier2007toward, smith2007web, post2000long}.

Collectively, legal systems typically comprise multiple heterogeneous components that interact in intricate ways, including various governing bodies, key participants, regulatory tools and even ethical principles, which form the basis on which law and politics aim to govern the complex web of values, perspectives and weights over competing priorities that are typical of human societies \cite{davis1970expository, stoetzer2015multidimensional, clark2010locating}.

Building upon the initial work \cite{vivo2024complexity, vivo2022physics, katz2020complex, leibon2018bending, ruhl2017harnessing, bommarito2017measuring, koniaris2018network, katz2010hustle, bourcier2007toward, smith2007web, post2000long}, the purpose of this Perspective paper is to highlight a selection of existing questions and issues in the legal domain that physicists and applied mathematicians might tackle using quantitative tools and techniques of complexity science. To begin, we will highlight some typical properties of ``complex'' legal systems and the wide variety of formal methods devised to tackle their underlying structure and dynamics.

\section{Properties of Legal Systems}

Some of the common properties present in many legal systems that contribute to the ``complexity'' dimension are summarised in the table below.  

\begin{table}[htpb]
  \centering
  \small
  \begin{tabular}{lc}
    \textbf{Common Properties of Legal Systems} 
    \\
        \hline
   \\
\textsc{Sources of Legal Rules}   \\
\textsc{Internal Actors \& Institutions}  \\
\textsc{External Actors \& Institutions}  \\
\textsc{Decisions \& Outputs}   \\
\textsc{Implicit \& Explicit Feedback Systems } \\ 
  &   \\
\hline
  \end{tabular}
  \label{tab:Properties}
\end{table}\textbf{}

\vspace{-.5cm}

\subsection{Sources of Legal Rules}
Modern legal systems create large bodies of written rules influencing the behaviour of individuals and institutions.  Source of rules can include narrow municipal law, grand sweeping legislation passed by a national legislature, and in certain instances can contain laws promulgated by super-national bodies such as the European Union.  Legal rules can also emanate from regulatory authorities who produce regulations that explicate the application of existing laws to various factual and concrete scenarios. Moreover, both judge-made rules in common law systems and the intricate jurisdictional protocols and processes associated with judicial decision making can also play an important role in the production and shaping of substantive rules \cite{macmahon2012proceduralism}. In all, this growing web of interconnected sources of law \cite{katz2020complex, coupette2021measuring} can be difficult for both lay end users and professional practitioners to easily decipher and navigate,\footnote{The U.K. Office of the Parliamentary Counsel (OPC) - a team of government lawyers who specialise in drafting legislation - launched the ``Good law'' initiative in 2013 \cite{GoodLaw} ``with a shared objective of making legislation work well for the users of today and tomorrow''. Notably, there are perhaps $50,000,000$ words of law currently in force in the U.K. - even if nobody knows for sure - and $\sim 100,000$ new words of law are produced every month \cite{Heaton}.} leading to practical and theoretical issues like internal inconsistencies and contradictions in the law and increased costs for the society \cite{ruhl2015measuring, katz2014measuring, de2022drafting}.

\subsection{Internal Actors \& Institutions}
Abstract legal rules must be interpreted and applied to the facts and circumstances of a given case.  Does a given rule implicate the particular conduct in question?  The initial determination of that question falls to a designated actor operating \emph{internally} within a given legal system, be it a law enforcement officer, a regulator, or ultimately one or more members of the judiciary. The internal psychology of these decision makers \cite{klein2010psychology, guthrie2000inside} and the institutional conditions surrounding the choices they make \cite{baum2009puzzle, segal2002supreme, epstein1997choices} -- including the famous statement that `justice is what the judge ate for breakfast' \cite{pnasextraneous} -- all contribute to a certain level of unwanted indeterminacy and ``fuzziness'' of legal outcomes and in turn to increased complexity. Among the most challenging issues facing modern societies and the handling of controversies lies also the technical and scientific competence (or lack thereof) of the internal agents, and their ability to correctly apply and understand the scientific method in the assessment of evidence and the balancing of arguments in concrete cases \cite{illiteracy1,illiteracy2,illiteracy3}.

\subsection{External Actors \& Institutions}
Legal systems do not operate in a vacuum and are not merely the byproduct of some sort of technocratic rule optimisation problem.  Rather, they govern the behaviour of individuals and organisations that in turn seek to influence the rule making and rule interpretation processes \cite{bombardini2020empirical, de2002lobbying} via public opinion movements, and lobbying from business leaders, trade groups, and other influential organisations. This \emph{co-evolutionary} perspective -- characterised by feedback and bottom-up pressure by the people (e.g. towards their elected representatives) is important for complexity scientists seeking to understand the behaviour of a legal system over time.  
\subsection{Decisions \& Outputs}
Legal systems decide things, and attempt to resolve not only broad societal controversies but also more mundane disputes between individual parties.  One question which looms over such decisions is the extent to which actors are actually constrained by legal rules. Namely, an often debated and complex question is the extent to which judges engage in post hoc rationalisation for their pure policy preferences as opposed to simply following existing precedents or statutory rules \cite{segal2002supreme, sunstein2007judges}.  To evaluate such questions, researchers often study decisions made by actors with decisions serving as a dependent variable whose outcomes are a function of a set of inputs including but not limited to the relevant set of legal rules \cite{segal2002supreme, pnasextraneous, sunstein2007judges}.

\subsection{Implicit \& Explicit Feedback Systems}
All of these constitutive elements are linked -- horizontally and vertically -- through procedures and mechanisms involving a non-trivial degree of \emph{randomness} such as legal proceedings, diplomatic discussions, and policy formulation processes. These mechanisms incorporate \emph{feedback loops}, including the direct appeals process and the interaction between public opinion and the outputs of legal and political systems \cite{hathaway2000path, jervis1998system, ruhl2015measuring}. Legal and political systems often exhibit \emph{self-regulating} characteristics, as seen in the evolution of judicial precedents or the codification of laws, whose dynamics can be \emph{non-linear}\cite{ruhl2017harnessing, smith2007web, post2000long, leicht2007large} and highly contingent on specific geographic and historic conditions. The individuals within this system typically operate under limited cognitive capacity and incomplete information and control over its dynamics.

\section{Formal Methods to Evaluate Legal Systems}

There are various methodological techniques that scholars have used to measure and model the behaviour and dynamics of legal systems.  As summarised in the table below, these approaches operate at different scales (macro/micro) and focus on different units of analysis (legal rules, actors, and outcomes).  For completeness, we include not only traditional complexity science methods but also a range of other quantitative, mathematical, or computational techniques.

\begin{table}[htpb]
  \centering
  \small
  \begin{tabular}{lc}
    \textbf{Formal Methods to Evaluate Legal Systems} 
    \\
        \hline
   \\
\textsc{Complex Networks}   \\
\textsc{Stat-Phys Models}  \\
\textsc{Game Theoretic Models} \\
\textsc{Empirical Evaluation of Legal Rules}   \\
\textsc{NLP \& Machine Learning Models} \\ 
  &   \\
\hline
  \end{tabular}
  \label{tab:Properties}
\end{table}\textbf{}

\vspace{-.5cm}

\subsection{Complex Networks} Complex networks have been widely used to represent interactions between agents/constituents in legal settings.
Two prominent incarnations come immediately to mind, both inspired by the seminal work of \cite{22}: (i) \emph{information networks} -- notably judicial citation networks \cite{leicht2007large,20,23,24,26} or legislation networks \cite{katz2020complex,28} -- where judicial decisions or pieces of legislation are connected if one cites or refers to the other giving rise to a hierarchy of \emph{precedents} that may be legally binding, and (ii) \emph{social networks} formed among agents operating in a legal environment, such as judges sitting on the same Court, or Courts dealing with the same case(s), or again members of Parliament or committees working on the same issue \cite{katz2010hustle,31}. Past research has focused on individual countries \cite{33,34,36} or comparisons between several countries \cite{37}, while there are also studies of networks at the European \cite{39,40,41,42} or the international level \cite{43,44,45,46}. 
By studying the citations between court decisions, researchers can potentially discern the content of cases, the connections between them, evolving trends in case law, and identify landmark cases that occupy central positions in the network \cite{CitNetwork1,CitNetwork5,CitNetwork6,CitNetwork8,CitNetwork9,CitNetwork10}. Similarly, patents are legal documents that require an inventor to situate their new innovation in relation to all prior patented inventions.  These ``prior art'' references constitute a dynamic directed acyclic graph (d-dag), whose properties can be explored using various graph mining techniques \cite{Patent1,Patent2, erdi2013prediction, 26}. Mapping citations to already existing technology onto a network structure, it is possible to identify the most cited patents, the most innovative and the most cited companies (innovation engines), as well as social properties such as \emph{homophily} (the inclination to cite patents from the same country or in the same language) and \emph{transitivity} (the inclination to cite references’ references).

\begin{figure}
\includegraphics[width=0.49\textwidth]{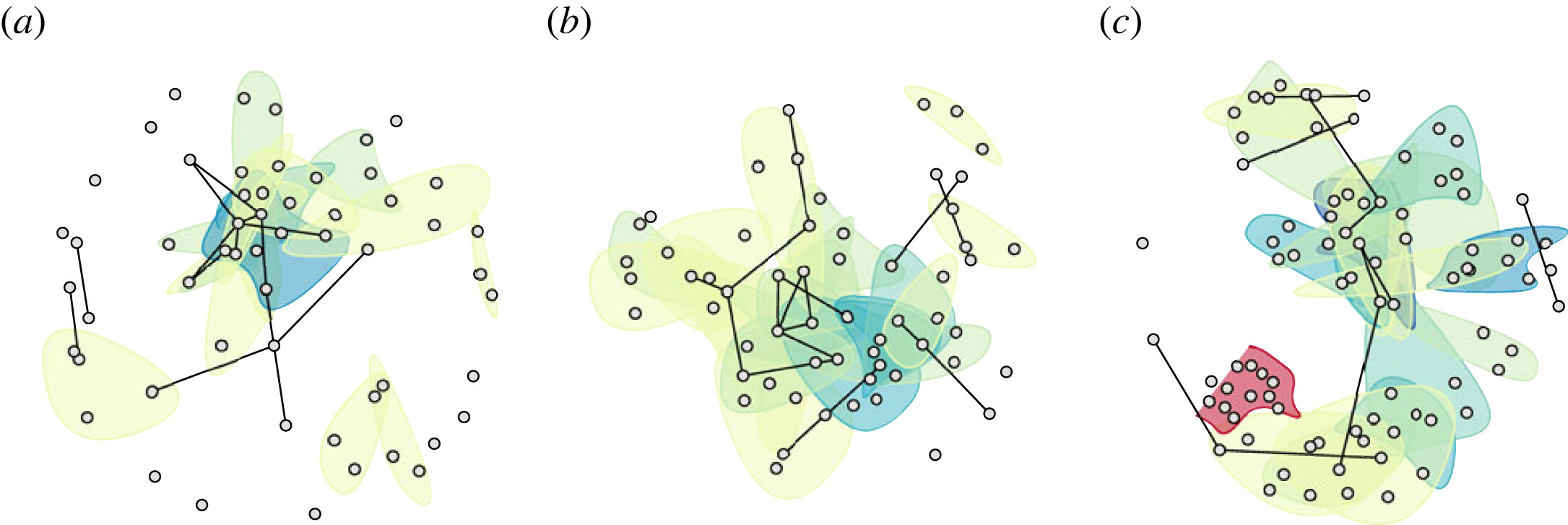}
\caption{\scriptsize{Three prominent decisions from Germany’s Federal Constitutional Court (GFCC) corpus depicted as hypergraphs (labeled volume, page). The decisions concern data privacy (a), European integration (b) and religious freedom (c), respectively. In each panel, nodes represent decisions cited at least once by the visualised decision, and (hyper)edges indicate unique citation blocks. Hyperedge colours progress, by increasing cardinality, from light yellow to dark blue and then red, with binary edges (indicating that exactly two decisions share a citation block) drawn in black. Figure taken from \cite{coupettehyper} and reproduced by virtue of CC BY license.}}
\label{hypergraph}
\end{figure}

Notable in this sphere is also the introduction of \emph{temporal hypergraphs} as a powerful tool for studying legal citation and collaboration networks \cite{coupettehyper}. Temporal hypergraphs generalise static graphs by (i) allowing any number of nodes to participate in an edge \cite{ouvrard,hyper1,hyper2,hyper3,hyper4,49}, and (ii) permitting nodes or edges to be added, modified, or deleted in time \cite{holme2012temporal,47} (see Fig. \ref{hypergraph} for an example from \cite{coupettehyper}). 

More recently, complex network analysis has been employed in the emerging field of \emph{corruptomics}, a multi-disciplinary enterprise aimed to characterise and identify corruption schemes and cartels \cite{Corruption1}, with important implications on public policies. Social network analysis provides insights also in criminology, from the identification of latent clusters of terrorist groups \cite{terror} to the analysis of criminal groups \cite{criminal2}, including prediction of links with affiliates not yet discovered \cite{criminal3,criminal4}.

Network analysis has also proven useful to identify \emph{pivotal} components in complex settings like political voting, finance, and social media \cite{LeePivotal}, namely those few agents whose behaviour is most representative or indicative of the collective outcome of a process, as well as to represent rights and other legal relations between individual legal actors \cite{sichelman}. 

\begin{figure}[h]
\begin{centering}
\includegraphics[width=0.49\textwidth]{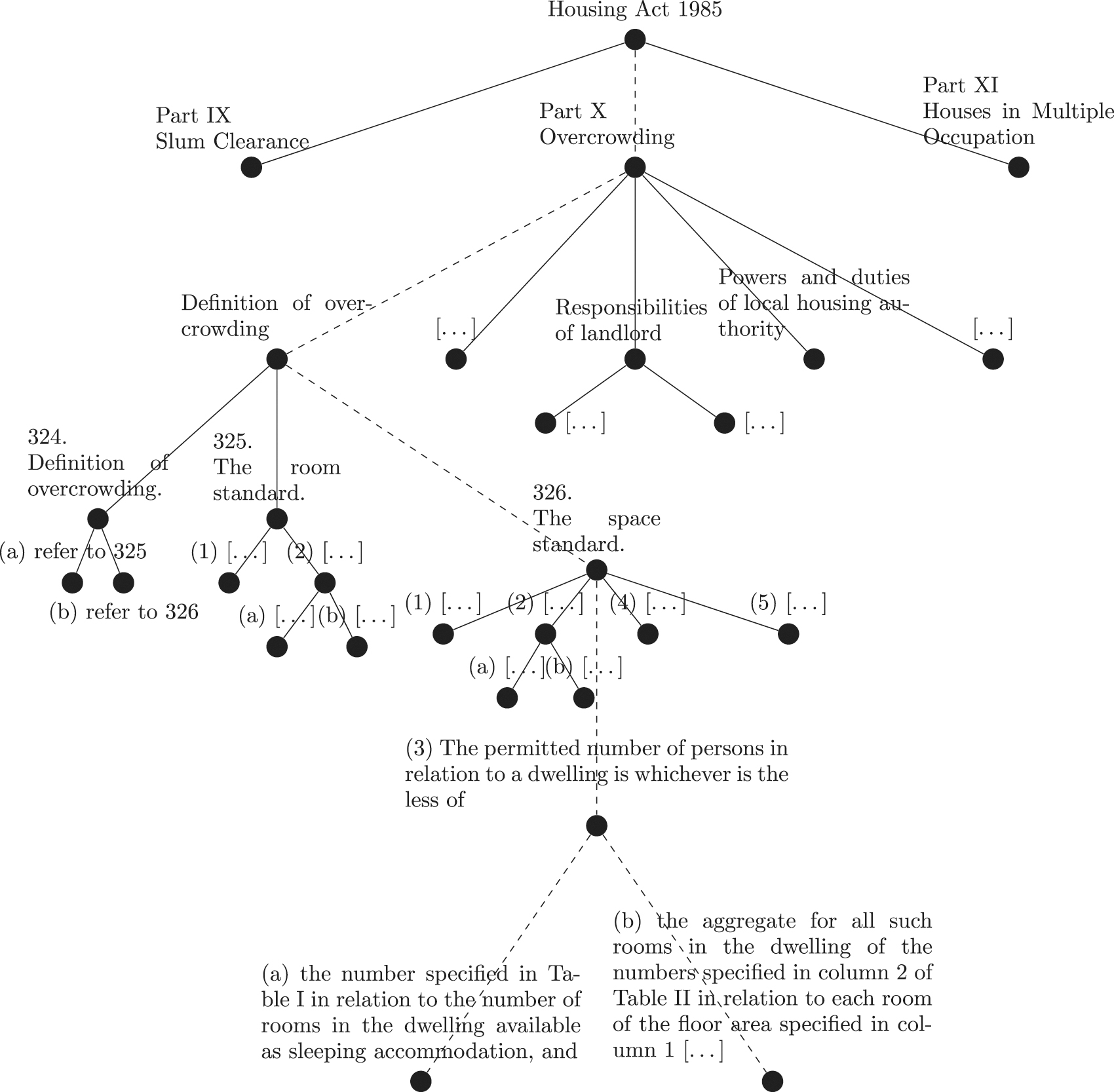}
\caption{\scriptsize{Example from an excerpt of the UK Housing Act 1985, c.68, to be found at \cite{UKHousing}. The nodes of the tree represent structural items of the text such as the Act itself, its Parts, Sections, etc, and two items
are linked if one is contained in the other. The dashed edges label the ideal path of a reader researching the question ‘What is the
maximum number of tenants that a UK landlord may let a property with a given number of rooms to without incurring in
penalties?’. Figure taken from \cite{retrieval} and reproduced by virtue of a CC BY license.}}
\label{overcrowding}
\end{centering}
\end{figure}

Another strand of research looks at network representation of legal texts (say, acts of parliaments), where legal provisions are organised in the form of a hierarchical tree (with Parts, Chapters, Paragraphs, Articles etc.), and different provisions within the same act (or across different acts) are further linked if they are connected via a citation or amendment. Network representations of legal texts allow us to define structural complexity metrics, for instance based on the mean time a `random reader' may take to retrieve a piece of information (see Fig. \ref{overcrowding} for an example) planted on a leaf node of the tree (see \cite{retrieval} and references therein) as well as to provide improved visualisation tools for legal texts that offer a better user experience allowing faster browsing and more efficient search capabilities \cite{CitNetwork10,visual0,visual1}.

Political science has also benefitted from a ``complexity science'' approach to e.g. representative democracy and its stability, including policy recommendations \cite{political2}. Among the most interesting questions in the political domain that can be tackled using complex network tools we mention the problem of optimal size of parliament of democratic countries, which are empirically known to scale as $\sim N^\alpha$ with the country population $N$, with an exponent $\alpha$ between $1/3$ and $1/2$ -- see \cite{gamberi2021maximal,taagepera1972size,margaritondo2021size} for theoretical models and quantitative predictions of the exponent $\alpha$.  

\subsection{Stat-Phys models} 
\begin{figure}
\begin{centering}
\includegraphics[width=0.49\textwidth]{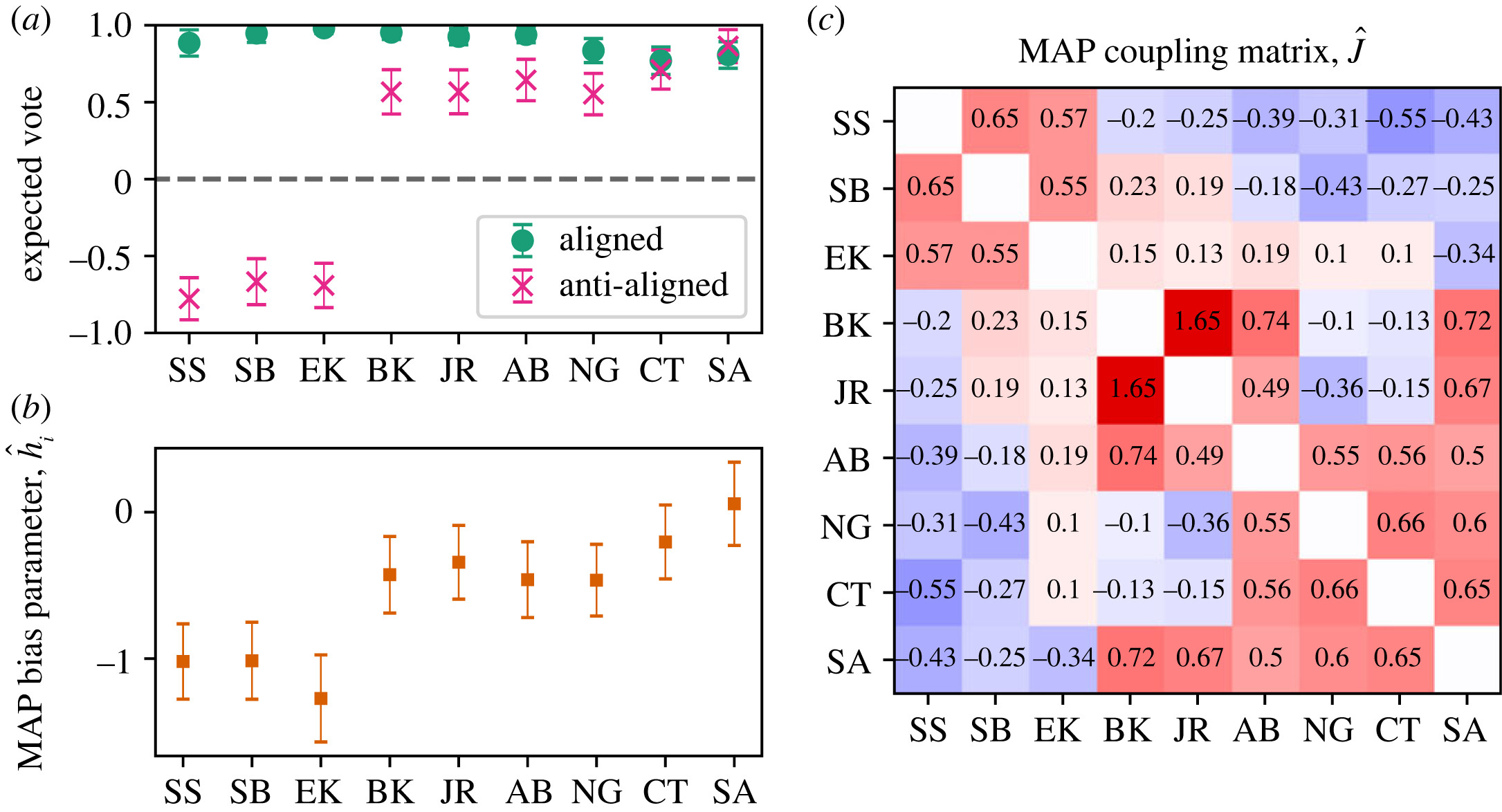}
\caption{\scriptsize{Model fits for the Roberts Court. (a) The expected vote of each judge, when the signs of \textit{H}
 and \textit{h} are aligned and when they are not. This corresponds to a near-unanimous behaviour and a 6-3 split. (b) Inferred most-likely values for the bias parameters, \textbf{\textit{h}}. (c) Inferred most-likely values for the couplings, \textit{J}. Interestingly, the strongest coupling by far is between Kavanaugh and Roberts. Judge initials SS for Sonia Sotomayor; SB, Stephen Breyer; EK, Elena Kagan; BK, Brett Kavanaugh; JR, John Roberts; AB, Amy Barrett; NG, Neil Gorsuch; CT, Clarence Thomas and SA, Samuel Alito. Error bars show one standard deviation in the posterior distribution. Figure taken from \cite{lee2024valence} and reproduced by virtue of a CC BY license.}}
\label{RobertsCourt}
\end{centering}
\end{figure}

More generally, models and concepts borrowed from classical Statistical Physics have been used in the legal context. In \cite{lee2015scotus,lee2024valence}, the opinions cast by the 9 US Supreme Court Justices (encoded in a vector $\bm\sigma$ of binary spin variables $\sigma_i=\pm 1$, with $i=1,\ldots,9$) and their empirically measured correlations $C_{ij}=\langle\sigma_i\sigma_j\rangle$ (i.e. justices casting their votes on the same or opposite sides) over the span of many years are used to calibrate a Maximum Entropy model yielding a Boltzmann-like distribution $P(\bm\sigma)=\exp(-E(\bm\sigma))/Z$. Here,  $E(\bm\sigma)=-(1/2)\sum_{i\neq j}J_{ij}\sigma_i\sigma_j$, with the ``Curie-Weiss'' couplings $J_{ij}$ tuned to reproduce the empirically observed correlations. Considering also the empirically measured expectation values of the votes from individual justices $\langle\sigma_i\rangle$, the energy function $E(\bm\sigma)$ can be corrected including an ``external field'' term $-\sum_i h_i\sigma_i$ accounting for external ideological biases. The model allows us to predict the probability of specific outcomes (say, $5-4$ on particularly contentious issues) as well as to dispel some myths about the ``political'' root of some decisions\footnote{Contrary to popular beliefs, roughly $\sim 50\%$ of US Supreme Court decisions are unanimous, and less than $\sim 20\%$ are split across the ``obvious'' ideological lines \cite{scotusdecision}.} (see Fig. \ref{RobertsCourt} for results from \cite{lee2024valence} for the Roberts Court). 

The classical concept of (informational) \emph{entropy} has found its way in legal studies as well. In \cite{StatPhys2} the language of legal codes from different countries and legal traditions was studied from the point of view of vocabulary entropy, which measures the diversity of the author’s choice of words, in combination with the compression factor, linked to the redundancy present in a text. In \cite{StatPhys3}, a quantitative formalisation of \emph{entropy} in the legal context is proposed, where it is argued that much of the ``work'' performed by the legal system is to reduce legal uncertainty by delineating, interpreting, and applying the law, a process that can in principle be quantified. The introduction of a ``randomised'' selection process for elected representatives in a large democratic house has proven beneficial in terms of quality and quantity of bills passed \cite{pluchino2011accidental}, a finding in line with the observed \emph{slackness} of many complex systems, which tend to work more efficiently and flexibly when some (small) sub-part does not strictly follow the overarching rules but rather adds noise and uncertainty \cite{pluchino2010peter,baek2009}. Classical Stat-Phys models such as the Potts model \cite{wu1982potts} have proven useful in the \emph{redistricting} problem \cite{chou2006taming}, namely the procedure by which the political power draws electoral districts to ensure that certain constitutional or statutory prescriptions are fulfilled --  including its distortions (e.g. \emph{gerrymandering}), where the districts are drawn to ensure maximal political advantage for the incumbent administration \cite{chou2006taming}.

\begin{figure}[h]
\begin{centering}
\includegraphics[width=0.4\textwidth]{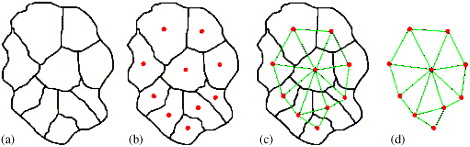}
\caption{\scriptsize{Network of voting precincts as developed in \cite{chou2006taming} (a) A district of 10 precincts in their model; (b) each red dot represents one precinct; (c) a network of precincts connected by green arcs; and (d) the underlying network extracted. Figure reproduced from \cite{chou2006taming} with permission from Elsevier.}}
\label{gerrymander}
\end{centering}
\end{figure}

\subsection{Game theoretic models} Lawyers and judges operate in strategic environments where the tools of game theory can help formalise how individuals craft strategies and make decisions \cite{baird1998game}. Thus, it is perhaps not surprising that game-theoretical concepts and frameworks have been adopted and translated into the legal domain quite early on.  While the vast majority of this work focused upon the application of simple ``prisoner's dilemma''-style models \cite{mcadams2008beyond}, there has also been more sophisticated work applying game theoretic models to a range of legal topics including tort law \cite{landes1980positive, shavell1980analysis, polinsky1991decoupling}, contract law \cite{avraham2006incomplete, hart1999foundations, ayres1989filling}, the selection of disputes in civil litigation \cite{priest1984selection, lee2016priest, klerman2018litigation} and the overall law making process \cite{tsebelis1999veto}. In addition, scholars have attempted to model how judges and courts operate in their own strategic environment. Early applications of game theory to judicial behaviour concentrated on the interactions between courts and the other branches of power \cite{vanberg2001legislative, rogers2001information}. More recently, the focus has shifted on characterising how judges on the same court may engage in `bargaining' strategies with each others to push the final outcome as close as possible to their ideal position \cite{lax2007bargaining, dyevre2011game}, while deciding a case on narrower grounds if this may help reach a wider consensus.  Other works have modelled the interactions among judges sitting on higher and lower courts and the degree to which lower courts may conform to the doctrine of precedent within the respective hierarchical structure \cite{de2002informative, kastellec2007panel}. Finally, the marriage between game theory and optimal stopping methods has achieved some progress in modelling the process of jury selection for criminal and civil trials in Common Law jurisdictions, and the strategic use of \emph{peremptory challenges} by Defense and Prosecution to strike down a pre-determined number of prospective jurors without any compulsory legal ground \cite{jury1,jury2}.

\subsection{Empirical Evaluation of Legal Rules}  
Legal rules, of course, have real world consequences for individuals and organisations.  As such, scholars have become increasingly interested in understanding whether legal rules actually achieved their desired ends.  For example, does the presence of a death penalty statute lead to fewer murders?  What are impacts of various corporate governance rules on long term firm performance?  Which types of intellectual property rules favour innovation?  Over the past several decades, the \textit{credibility} revolution which swept through empirical economics \cite{angrist2010credibility} has also made its way into legal studies \cite{ho2011credible}. In a manner akin to evidence based medicine, the use of empirical models and methods has increasingly become part of mainstream legal scholarship.

The dynamics underlying many phenomena are complex and the underlying complexity can make it difficult to disentangle cause and effect \cite{sobel2000causal}.  Econometricians have developed a number of sophisticated techniques to help tease out core underlying causal relationships \cite{sobel2000causal} and these techniques are on display in a wide range of leading empirically oriented legal studies \cite{ho2011credible}. However, there is still plenty of opportunity for future work.  

Namely, linking cause and effect from retrospective data snapshots does not fully address the challenge of generating robust forward predictions (particularly in non-stationary environments) \cite{breiman2001statistical, barrett2006prediction}.  Ultimately, it is these forward predictions that are important to guide policy makers as they help shape empirically informed public policy. 

\subsection{NLP \& Machine Learning Models}  
The complexity of legal language is challenging for both laypersons and experts alike \cite{StatPhys2, katz2014measuring, martinez2023even}. Indeed, members of the public often use terms such as ‘legalese’ and ‘legal gobbledygook’ to characterise the specialised vocabulary and intricate concepts contained in legal language. To better understand the underlying dynamics and to help support the more thorough exploration of various legal systems, a range of scholars studying legal systems have utilised methods from Natural Language Processing (NLP) and machine learning \cite{frankenreiter2020computational}. Such methods have also been leveraged to help support both compliance with existing legal rules \cite{greenhill2024machine} and to improve users ability to access legal information \cite{CaselawAccess, schwartz2024}.

Most recently, Large Language Models \cite{min2023recent} -- pretrained transformer-based models optimised for purpose-built, distributed GPU/TPU infrastructure -- have already demonstrated significant initial promise on both research and substantive tasks \cite{dell2023navigating, katz2024gpt}.  Notwithstanding, it is important that oversight and caution be paid as there is still the potential for models to hallucinate. At the same time, the performance of such foundational models will likely improve particularly when connected to additional engineering enhancements such as agentic capabilities. 

\section{Conclusion} In this Perspective, we have provided an overview of how concepts and methods from complexity science and statistical physics can be applied to study legal systems. We have highlighted various approaches, ranging from network analysis of legal citations to game theoretic models of judicial decision-making and empirical evaluations of legal rules. These diverse methods offer powerful tools to analyse the intricate web of interactions, feedback loops, and emergent behaviours that characterise modern legal systems. 
Looking ahead, several promising avenues for future research emerge. First, there is significant potential to leverage Large Language Models to analyse vast corpora of legal texts at scale, potentially uncovering hidden patterns and relationships. Second, the development of more sophisticated agent-based models incorporating realistic cognitive and institutional constraints could shed light on the emergence of legal norms, the impact of policy interventions, and the much debated problem of ``automated'' (AI-assisted) trial decisions that forego (or make less relevant) the ``human'' factor. Interdisciplinary collaborations bringing together legal scholars, complexity scientists, and domain experts and practitioners will be crucial to ensure that quantitative models remain grounded in legal reality and adhere to long-standing constitutional and statutory principles.

\acknowledgments
P.V. acknowledges support from UKRI Future Leaders Fellowship Scheme (No. MR/X023028/1).

\end{document}